\documentstyle[12pt]{article}
\begin{document}

\title{Enthalpy recovery in semicrystalline polymers}

\author{Aleksey D. Drozdov\\
Institute for Industrial Mathematics\\
4 Hanachtom Street, Beersheba, 84311 Israel}
\date{}
\maketitle

\begin{abstract}
Constitutive equations are derived for enthalpy recovery in polymeric
glasses after thermal jumps.
The model is based on the theory of cooperative relaxation in a
version of the trapping concept.
It is demonstrated that some critical temperature $T_{\rm cr}$
and some critical degree of crystallinity $f_{\rm cr}$ exist
in a semicrystalline polymer above which
the energy landscape becomes homogeneous and structural relaxation ceases.
\end{abstract}

\section{Introduction}

This note is concerned with the kinetics of enthalpy recovery
in polymeric glasses after thermal jumps.
Physical aging of polymers has been the focus of attention
in the past three decades,
see monographs \cite{Str78,Bra85,Don92,Mat92,DB96}
and review articles \cite{MK89,MCO91,EAN96}.
Despite significant successes in the study of out-of-equilibrium
dynamics in disordered materials achieved in the past decade
(see reviews \cite{Kob97,BCK98,Bou99,HVD99,KBS00,TK00}),
it is difficult to mention a model that adequately predicts
structural relaxation in polymers.
Even the mode-coupling theory \cite{GS92}
(the most advanced among molecular models)
fails to describe slowing down in the response of amorphous
polymers below the glass transition temperature $T_{\rm g}$
\cite{KBS00,Dyr95,CLH97}.
As a reason for this conclusion, the neglect of cooperativity in
the molecular reorientation is conventionally mentioned
\cite{AG65,MH97}.

Enthalpy relaxation provides one of the most convenient ways to study
aging of glassy polymers \cite{MCO91,Hod94,RP95,HSH96,BCF97}.
The kinetics of enthalpy recovery is in close connections
with time-varying changes in static and dynamic moduli
\cite{BF95,Yos95,CFH98},
specific volume \cite{CHM98,Dro99}
and yield stress \cite{BCB82,CME99,HSH99,PGM00},
as well as with transition from ductile to brittle
fracture \cite{JL92,MJ95}.
For some polymers (e.g., polyesters), DSC measurements demonstrate
a dramatic effect of the waiting time \cite{AH82},
whereas the free volume fraction determined by PALS shows
no pronounced decrease induced by physical aging \cite{MDJ00}.

Constitutive models for structural relaxation in glassy polymers
are conventionally confined to amorphous media (as exceptions,
we would mention Refs. \cite{CM80,Str87}).
This may be explained by a belief that crystalline lamellas
do not affect the process of structural relaxation occurring
in amorphous regions \cite{TW81}.
This hypothesis may be accepted provided that only one
kind of amorphous domains exists in a semicrystalline polymer.
A number of observations demonstrate, however, a more complicated picture,
where two different kinds of amorphous regions co-exist in a partially
crystallized polymer: amorphous layers between lamellas
and amorphous domains between spherulites \cite{IB63,GBS76,LC96,DCM99}.
This results in a question whether the rate of physical aging is a linear
function of the degree of crystallinity which vanishes when the degree
of crystallization $f$ reaches 100\% only (as it is postulated in Fig.~6
of Ref. \cite{TW81}), or the rates of structural relaxation substantially
differ in interlamellar and intralamellar zones which implies that
physical aging may decline even in a partially crystallized material.

The objective of this study is to derive a constitutive model
for enthalpy recovery which is based on the theory of cooperative
relaxation in a version of the trapping concept
\cite{Dyr95,Bou92,MB96,Sol98,DMO99}.
To simplify the analysis, we do not distinguish explicitly between
interlamellar and intralamellar regions, but presume that the ``averaged''
distribution of energies of traps may depend on the degree of crystallinity.
An amorphous region is treated as an ensemble of mutually independent
cooperatively rearranged regions (CRR).
A CRR is thought of as a globule consisting of scores of strands of long
chains \cite{Sol98}.
The characteristic length of a CRR in the vicinity of $T_{\rm g}$
amounts to several nanometers \cite{RN99}.
In the phase space, a CRR is treated as a point trapped in its
potential well (cage).
At random times, the unit hops to higher energy levels as it is
thermally agitated.
When the energy of thermal fluctuation exceeds the height of a barrier
between cages, a CRR can change its trap.
Introducing several hypotheses regarding the kinetics of hops from
one potential well to another,
we develop a nonlinear parabolic equation for the probability density
of traps with various potential energies.
This equation is applied to the analysis of enthalpy recovery
in amorphous and semicrystalline poly(ethylene terephthalate) (PET).
The following conclusions are drawn:
\begin{enumerate}
\item
the degree of crystallinity of a semicrystalline polymer
affects the model's parameters in a similar way as temperature
influences those for an amorphous polymer
(a crystallinity--temperature principle analogous to
the time--temperature principle in linear viscoelasticity \cite{Fer80});

\item
a critical level of crystallinity $f_{\rm cr}<100$\% exists
at which the energy landscape becomes homogeneous.
This level may be treated as an analog to the critical temperature
for structural relaxation in amorphous polymers \cite{Dro00}
predicted by the mode-coupling theory.
\end{enumerate}
The exposition is organized as follows.
Constitutive equations for the kinetics of enthalpy relaxation are
developed in Section 2.
These relations are verified in Section 3 by comparison with
experimental data.
Some concluding remarks are formulated in Section 4.

\section{A model for enthalpy relaxation}

Denote by $w$ the energy of a potential well with respect
to some reference state \cite{Gol69}.
It is assumed that $w>0$ for any trap and $w=0$ for the reference state.
At random times, a CRR hops to higher energy levels in its potential well
as it is thermally activated.
Denote by $q(\omega)d\omega$ the probability to reach (in a hop)
the energy level that exceeds the bottom level of its potential well by
a value located in the interval $[\omega, \omega+d\omega ]$.
Referring to \cite{BCK98}, we set
$q(\omega)=A\exp(-A\omega)$, where $A$ is a material constant.
The probability for a CRR in a trap with potential energy $w$
to reach the reference state in an arbitrary hop is given by
\[ Q(w)=\int_{w}^{\infty} q(\omega)d\omega =\exp( -A w). \]
The average rate of hops in a cage $\gamma$ is determined by
the current temperature $T$ only, $\gamma=\gamma(T)$.
The rate of rearrangement $P$ equals the product of the rate
of hops $\gamma$ by the probability $Q$ to reach the reference
state in a hop,
\begin{equation}
P(w)=\gamma \exp(-Aw).
\end{equation}
Denote by $X$ the (time-uniform) concentration of traps per unit mass,
and by $p(t,w)$ the current probability density of traps with potential
energy $w$.
The number of relaxing regions (per unit mass) trapped in cages with
the energy belonging to the interval $[w, w+dw]$
and rearranged during the interval of time $[t,t+dt]$ is $X P(w)p(t,w)dwdt$.
Unlike Refs. \cite{Rob78,RSC84},
we assume that not all flow units change their traps when they
reach the reference energy level and denote by $F(t,w)$ the ratio
of the number of relaxing regions returning to their traps to the
number of those reaching the reference state.
The number of relaxing regions leaving their cages
(with the energy located within the interval $[w,w+dw]$) per unit
mass and unit time is given by $X (1-F) P p dw$.
The exchange of flow units is assumed to occur only between the nearest
neighbors on the energy landscape, that is between a trap with
the energy $[w,w+dw]$ and traps with the energies $[w-dw,w]$ and
$[w+dw,w+2dw]$.
The balance law for the number of flow units trapped in cages
with the energy belonging to the interval $[w,w+dw]$ reads
\[ \frac{\partial p}{\partial t}=-(1-F)Pp
+\frac{1}{2}\Bigl [(1-F)Pp\Bigr ]_{+}
+\frac{1}{2}\Bigl [(1-F)Pp\Bigr ]_{-}, \]
where the subscript indices ``$-$'' and ``$+$'' refer to appropriate
quantities for the intervals $[w-dw,w]$ and $[w+dw,w+2dw]$.
Expanding the right-hand side of this equality into the Taylor series,
using Eq. (1) and introducing the notation
$\Gamma=\frac{1}{2}\gamma dw^{2}$,
we arrive at the differential equation for diffusion over the energy
landscape
\begin{equation}
\frac{\partial p}{\partial t}=\Gamma \frac{\partial^{2}}{\partial w^{2}}
\Bigl [(1-F)\exp (-Aw)p\Bigr ].
\end{equation}
We adopt the Metropolis transition rates \cite{BB85},
\begin{equation}
F(t,w)=\left \{ \begin{array}{ll}
1, & p(t,w)\leq p_{\infty} (w),\\
\exp [-\epsilon(p-p_{\infty})], & p(t,w)>p_{\infty}(w),
\end{array}\right .
\end{equation}
where $p_{\infty}(w)$ is the equilibrium density of traps
and $\epsilon>0$ is a material parameter.
An important advantage of Eq. (2) compared to relationships suggested
in Refs. \cite{Fel68,Dob80} is that under condition (3) it does not
impose restrictions on the equilibrium density of trap $p_{\infty}(w)$.
Referring to \cite{Dyr95}, we suppose that the inequality
\begin{equation}
\int_{-\infty}^{0} p(t,w)dw\ll 1
\end{equation}
is satisfied for any $t\geq 0$
and describe the initial distribution, $p_{0}(w)$,
and the equilibrium distribution, $p_{\infty}(w)$,
by the Gaussian formulas
\begin{equation}
p_{0}(w) = \frac{1}{\sqrt{2\pi}\Sigma_{0}}\exp \biggl [
-\frac{(w-W)^{2}}{2\Sigma_{0}^{2}}\biggr ],
\qquad
p_{\infty}(w) = \frac{1}{\sqrt{2\pi}\Sigma_{\infty}}\exp \biggl [
-\frac{(w-W)^{2}}{2\Sigma_{\infty}^{2}}\biggr ],
\end{equation}
where $W$, $\Sigma_{0}$, $\Sigma_{\infty}$ are adjustable parameters.
Equations (5) imply that the average equilibrium energies of traps
are temperature-independent (the same value $W$ is employed for the initial
and equilibrium distribution functions), whereas their variances
strongly depend on $T$.
The first assertion is fairly well confirmed by experimental data
in mechanical tests \cite{Dro99a,Dro99b},
whereas the other hypothesis is in agreement with the conventional
scenario for the growth in the ruggedness of the energy landscape with
a decrease in temperature \cite{BCK98,DMO99,RSN99}.

The level of disorder in an ensemble of CRRs is characterized by
the configurational entropy per rearranging region \cite{AM88},
\[ s(t)=-k_{B}\int_{0}^{\infty} p(t,w) \ln p(t,w) dw, \]
where $k_{B}$ is Boltzmann's constant.
The configurational enthalpy per unit cage $h(t)$ is expressed in terms
of the configurational entropy $s(t)$ by means of the conventional
formula
\[ \frac{\partial h}{\partial s}=T. \]
This equality is integrated for a standard one-step thermal test,
\begin{equation}
T(t)=T_{0} \quad (T<0),
\qquad
T(t)=T \qquad (T>0),
\end{equation}
and an explicit formula is found for the enthalpy per unit mass $H=X h$.
The relaxing enthalpy per unit mass $\Delta H(t)=H(t)-H(0)$
is given by
\begin{equation}
\Delta H(t)= \Lambda \int_{0}^{\infty} \Bigl [ p_{0}(w)\ln p_{0}(w)
-p(t,w)\ln p(t,w) \Bigr ] dw
\end{equation}
with $\Lambda=k_{B}TX$.
Introducing the dimensionless variables $\bar{w}=Aw$ and $\bar{t}=t/t_{0}$,
where $t_{0}$ is the characteristic time of aging,
and setting $\bar{\Gamma}=A^{2}\Gamma t_{0}$,
$\bar{W}=AW$ and $\bar{\Sigma}_{k}=A\Sigma_{k}$,
we arrive at the constitutive model, Eqs. (2), (3), (5) and (7),
with six adjustable parameters $\bar{W}$, $\bar{\Sigma}_{0}$,
$\bar{\Sigma}_{\infty}$, $\bar{\Gamma}$, $\epsilon$ and $\Lambda$.
These kinetic equations substantially differ from conventional relations
for enthalpy recovery in polymers \cite{RP95,MEW74,KAH79,Hod87},
because they do not refer to a (purely phenomenological) concept of
internal clock.
As an analog of the material time $\tau$ in Eqs. (2), (3), (5) and (7),
we may mention the parameter $\epsilon$ that characterizes
the rate of changes in the energy landscape.
An important difference between $\epsilon$ and $\tau$ is that
$\epsilon$ is independent of the current energy landscape
(but, in general, temperature-dependent),
whereas the parameter $\tau$ is conventionally expressed
in terms of the current enthalpy $H$ by means of the Narayanaswamy
or the Adam--Gibbs equations.

\section{Comparison with experiments}

It is easy to check that the quantities $\bar{W}$ and $\bar{\Gamma}$
are interrelated: when one of them is fixed, the other may be chosen
to characterize the time scale.
For convenience of numerical simulation, we fix $\bar{W}$
and determine $\bar{\Gamma}$ by fitting observations.
The value $\bar{W}=2.5$ ensures that inequality (4) is satisfied with
a high level of accuracy.

We begin with experimental data for two amorphous PETs
exposed in Ref. \cite{JL92}.
First, observations are fitted for a homopolymer
to determine the parameters $\bar{\Sigma}_{0}$,
$\bar{\Sigma}_{\infty}$, $\bar{\Gamma}$, $\epsilon$ and $\Lambda$
that ensure the best approximation of observations.
Afterwards, the amounts $\bar{\Sigma}_{0}$,
$\bar{\Sigma}_{\infty}$ and $\bar{\Gamma}$ are fixed,
and the quantities $\epsilon$ and $\Lambda$ are determined
by matching experimental data for a copolymer.
Figure~1A demonstrates fair agreement between observations and results
of numerical simulation.
Using the specific gravity $g=1.34$ g/cm$^{3}$ at $T=65$~$^{\circ}$C
\cite{JL92}, we find the volume concentration of relaxing regions in PET,
$\Xi=6.39\cdot 10^{26}$ m$^{-3}$,
which is rather close to $\Xi=4.1\cdot 10^{26}$ m$^{-3}$
for polycarbonate \cite{BWK99}
and $\Xi=7.7\cdot 10^{26}$ m$^{-3}$ for poly(vinyl acetate)
\cite{DLS00} found by PALS.
To ensure that adjustable parameters are rather robust with respect
to changes in material properties and conditions of the test (a decrease
in $T_{\rm g}$ by 5~K and an increase in $T_{0}$ by 15~K), we repeat
this procedure using experimental data presented in Ref. \cite{Pet74}.
Figure~1B demonstrates fair agreement between observations
and results of numerical simulation for similar values
of $\bar{\Sigma}_{0}$, $\epsilon$ and $\Lambda$.
An increase in $\bar{\Gamma}$ depicted in Figure~1B compared to Figure~1A
seems quite natural, because this parameter is determined by the
difference between the glass transition temperature and the annealing
temperature, $\Delta T=T_{\rm g}-T$.
A decrease in $\bar{\Sigma}_{\infty}$ may be associated with a drop
in $T_{\rm g}$, because it is conventionally accepted that the glass
transition temperature grows when molecular mobility becomes more
restricted (e.g., because of an increase in the number of
crosslinks \cite{CME99} or in the molecular weight \cite{AHP76}),
which may be associated with the growth of inhomogeneity of the energy
landscape.

To study the effect of the annealing temperature $T$ on the kinetics
of structural recovery, we approximate experimental data for
amorphous PET obtained in Ref. \cite{AH82}.
Figure~2A demonstrates fair agreement between observations and results
of numerical simulation in the temperature range from $T_{\rm g}-25$
to $T_{\rm g}-5$~$^{\circ}$C.
The standard deviation of energies of traps in thermodynamic
equilibrium $\bar{\Sigma}_{\infty}$ is plotted versus the degree of
undercooling $\Delta T$ in Figure~2B.
This figure demonstrate that the dependence $\bar{\Sigma}_{\infty}(T)$
is fairly well approximated by the linear function
\begin{equation}
\bar{\Sigma}_{\infty}=a_{0}+a_{1}\Delta T,
\end{equation}
where $a_{k}$ are adjustable parameters.
In follows from Eq. (8) that some critical temperature $T_{\rm cr}$
exists at which $\bar{\Sigma}_{\infty}$ vanishes and the energy
landscape becomes homogeneous.
Results of simulation imply that $T_{\rm cr}=T_{\rm g}+34.11$~$^{\circ}$C
which is in accord with the values of $T_{\rm cr}$ found by fitting
date in mechanical tests \cite{Dro00}.
The quantities $\bar{\Gamma}$ and $\epsilon$ are depicted in Figure~3A.
This figure show that the dependences $\bar{\Gamma}(T)$
and $\epsilon(T)$ are correctly approximated by the ``linear''
functions
\begin{equation}
\log \bar{\Gamma}=b_{0}-b_{1}\Delta T,
\qquad
\log \epsilon=c_{0}+c_{1}\Delta T
\end{equation}
with adjustable parameters $b_{k}$ and $c_{k}$.
In the vicinity of the glass transition temperature, the apparent
activation energy $\Delta E$ is calculated as \cite{Str97}
\[ \Delta E=-R\frac{d\ln \bar{\Gamma}}{d(1/T)} \biggl |_{T=T_{\rm g}}, \]
where $R$ is the gas constant.
This equality together with Eq. (9) results in the formula
$\Delta E=RT_{\rm g}^{2}b_{1}$.
According to Figure~3A, $\Delta E=156.1$ kJ/mol, which is quite comparable
with $\Delta E=206$ kJ/mol determined for PET in mechanical tests
\cite{CYM99}.
An increase in the parameter $\bar{\Gamma}$ with temperature $T$ is
in agreement with the theory of thermally activated processes.
A surprising result is a decrease in $\epsilon$ with temperature.
It follows from Eq. (3) that this decrease provides an additional source
for slowing down of the aging process in the close vicinity of $T_{\rm g}$
which has not been accounted for in previous studies.

To assess the effect of crystallinity on the rate of structural relaxation,
we approximate experimental data for enthalpy recovery in semicrystalline
PET with various degrees of crystallinity $f$.
For a detailed description of specimens and the experimental procedure,
see Ref. \cite{DCM99}.
First, we approximate observations for a sample with the smallest
value of $f$ and determine adjustable parameters of the model
using the steepest-descent procedure.
Afterwards, we fix the quantities $\bar{\Sigma}_{0}$ and $\Lambda$
and repeat matching experimental data with three adjustable parameters,
$\bar{\Sigma}_{\infty}$, $\bar{\Gamma}$ and $\epsilon$.
Figure~4A demonstrates fair agreement between observations and predictions
of the model.
The equilibrium standard deviation of energies of traps
$\bar{\Sigma}_{\infty}$ is depicted versus the degree of crystallinity $f$
in Figure~4B.
The dependence $\bar{\Sigma}_{\infty}(f)$ is correctly approximated
by the linear function
\begin{equation}
\bar{\Sigma}_{\infty}=a_{0}-a_{1}f
\end{equation}
with adjustable parameters $a_{k}$.
Comparing Figures~2B and 4B, we conclude that an increase in $f$ for
a semicrystalline polymer affects the equilibrium distribution of cages
in a way similar to that in which the annealing temperature $T$
influences the energy landscape for an amorphous material.
By analogy with an amorphous polymer, one may define the critical
degree of crystallinity $f_{\rm cr}$ for a semicrystalline
medium as the percentage of crystallites at which the parameter
$\bar{\Sigma}_{\infty}$ vanishes and the energy landscape becomes
homogeneous.
In contrast to the conclusions of Ref. \cite{TW81}, Figure~4B
demonstrates that $f_{\rm cr}$ is essentially less than 100\%.
The parameters $\bar{\Gamma}$ and $\epsilon$ are plotted versus
the degree of crystallinity in Figure~3B.
This figure shows that the dependences $\bar{\Gamma}(f)$
and $\epsilon(f)$ are correctly approximated by the ``linear''
functions
\begin{equation}
\log \bar{\Gamma}=b_{0}+b_{1}f,
\qquad
\log \epsilon=c_{0}-c_{1}f
\end{equation}
with adjustable parameters $b_{k}$ and $c_{k}$.

Two kinds of crystallization of polymers are conventionally studied:
thermal crystallization, when a specimen is annealed at a fixed
temperature in the rubbery region for a given time (this procedure
was employed in Ref. \cite{DCM99}) and mechanically induced
crystallization, when a sample is stretched slightly above $T_{\rm g}$
to a given extension ratio $\lambda$.
To demonstrate that the effect of the degree of crystallinity
on the kinetics of physical aging weakly depends on the crystallization
procedure, we approximate experimental data for semicrystalline PET
samples where the level of crystallization is established by uniaxial
stretching at an elevated temperature.
A detailed description of specimens and the experimental procedure
can be found in Ref. \cite{MJ95}.
The extension ratio $\lambda$ is transformed into the level
of crystallinity using Figure~1 of Ref. \cite{VLJ98}.
Unlike previous sets of experimental data, where $\epsilon$ is treated
as a function of the level of crystallinity $f$, we fit observations
at various elongations $\lambda$ with the same value of $\epsilon$
found for an undeformed specimen.
Figure~5A demonstrates good agreement between experimental data and
predictions of the model.
The functions $\bar{\Sigma}_{\infty}(f)$ and $\bar{\Gamma}(f)$ are plotted
in Figure~5B.
This figure demonstrates that Eqs. (10) and (11) adequately
describe observations.
The parameter $b_{1}$ (that characterizes the influence of crystallinity
on the rate of rearrangement) acquires similar values for thermally
and mechanically crystallized samples.
The same assertion is true for the critical degree of crystallinity
$f_{\rm cr}$ which adopts the value 33.9\% for the stretched PET
versus 42.7\% for the PET annealed above $T_{\rm g}$.
On the contrary, the parameter $\bar{\Sigma}_{\infty}$ for the PET obtained
by thermal crystallization exceeds that for the hot-drawn PET by twice.
This may be explain by the fact that stretching a rubbery polymer
establishes an additional order caused by partial destruction
of a polymeric network and alignment of long chains along the axis of
loading \cite{PGM00}.

\section{Concludung remarks}

Constitutive equations have been derived for enthalpy relaxation
in glassy polymers after thermal jumps.
The model is based on the trapping concept which treats a disordered
medium as an ensemble of flow units rearranged at random times
as they are thermally activated.
Adjustable parameters are found by fitting experimental data
for amorphous and semicrystalline poly(ethylene terephthalate).
The following conclusions are drawn:
\begin{enumerate}
\item
the constitutive equations correctly describe the kinetics
of structural relaxation in the sub--$T_{\rm g}$ region.

\item
the model predicts the existence of some critical
temperature $T_{\rm cr}$ at which the energy landscape
becomes homogeneous.

\item
for semicrystalline polymers, the model implies the existence of
some critical degree of crystallinity $f_{\rm cr}$ at which the
energy landscape becomes homogeneous.
The inequality $f_{\rm cr}<100$\% may serve as an indirect confirmation
that the processes of structural relaxation in interlamellar and
intralamellar regions differ from one another.

\item
a correspondence may be established between the effects of
temperature and the level of crystallinity on the kinetics of
enthalpy recovery.
\end{enumerate}

\subsection*{Acknowledgments}

The author is grateful to A.L. Svistkov for fruitful discussions.
The work is supported by the Israeli Ministry of Science through
grant 1202--1--98.

\newpage
\setlength{\unitlength}{0.5 mm}
\begin{figure}[t]
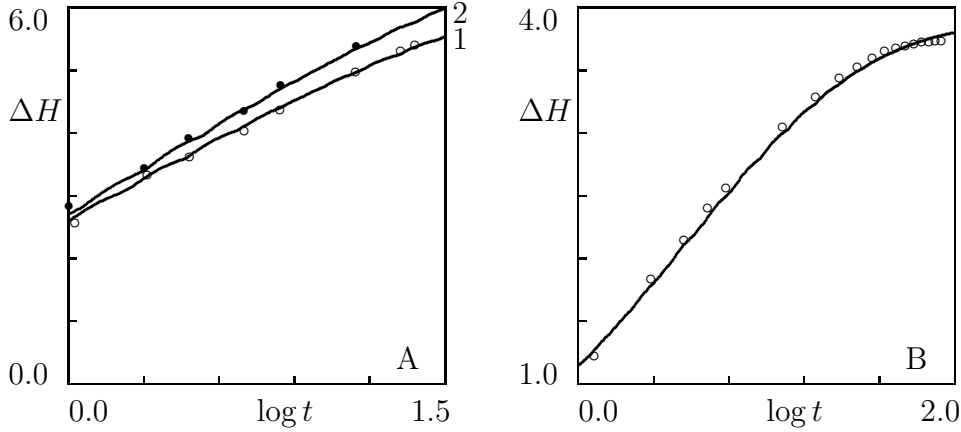

\begin{center}

\end{center}

\caption{A---The relaxation enthalpy $\Delta H$ J/g versus time $t$ days
for amorphous PET ($T_{\rm g}=80$~$^{\circ}$C)
quenched from $T_{0}=80$ to $T=65$~$^{\circ}$C.
Circles: experimental data \protect{\cite{JL92}}.
Solid lines: prediction of the model with
$\bar{\Sigma}_{0}=0.19$, $\bar{\Sigma}_{\infty}=0.95$ and
$\bar{\Gamma}=8.0$~days$^{-1}$.
Curve~1: homopolymer, $\epsilon=0.2$, $\Lambda=4.0$~J/g;
curve~2: copolymer (3.5 mole \% of cyclohexane dimethanol),
$\epsilon=0.17$, $\Lambda=4.4$~J/g.
B---The relaxation enthalpy $\Delta H$ J/g versus time $t$ h
for amorphous PET ($T_{\rm g}=75$~$^{\circ}$C)
quenched from $T_{0}=95$ to $T=65$~$^{\circ}$C.
Circles: experimental data \protect{\cite{Pet74}}.
Solid line: prediction of the model with
$\bar{\Sigma}_{0}=0.15$, $\bar{\Sigma}_{\infty}=0.38$,
$\bar{\Gamma}=1.24$~h$^{-1}$, $\epsilon=0.2$ and $\Lambda=4.2$~J/g}
\end{figure}

\newpage

\begin{figure}[t]
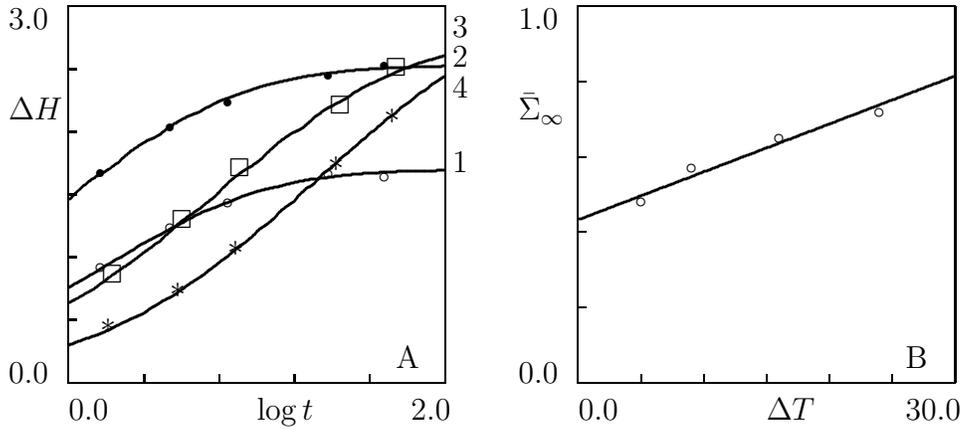

\begin{center}

\end{center}

\caption{A---The relaxation enthalpy $\Delta H$ J/g versus time $t$ h
for amorphous PET ($T_{\rm g}=76$~$^{\circ}$C)
quenched from $T_{0}=277$ to the test temperature $T$~$^{\circ}$C.
Symbols: experimental data \protect{\cite{AH82}}.
Solid lines: prediction of the model with $\bar{\Sigma}_{0}=0.25$
and $\Lambda(T_{\rm g})=2.9$~J/g.
Curve~1: $T=71$;
curve~2: $T=67$;
curve~3: $T=60$;
curve~4: $T=52$.
B---The parameter $\bar{\Sigma}_{\infty}$ versus $\Delta T$~$^{\circ}$C.
Circles: treatment of observations.
Solid line: approximation of the experimental data by Eq. (8)
with $a_{0}=0.433$ and $a_{1}=0.013$}
\end{figure}

\begin{figure}[t]
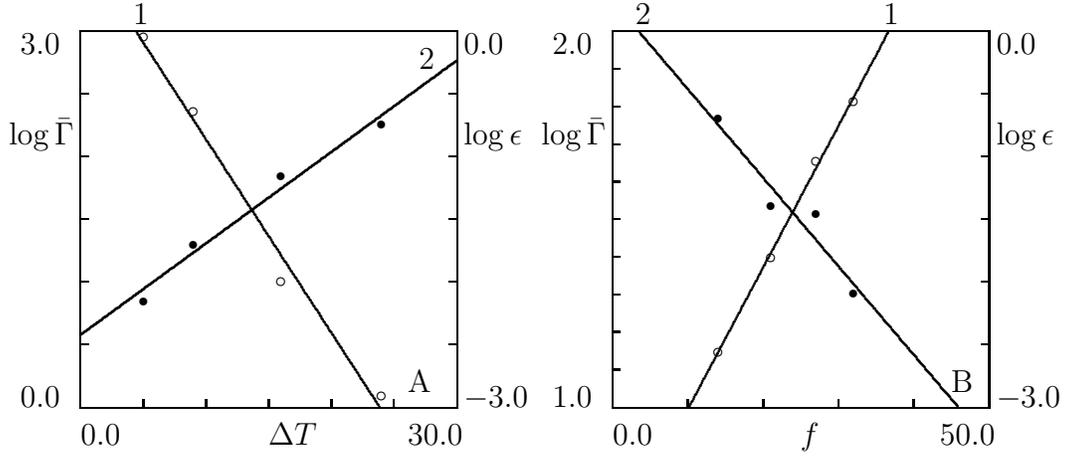

\begin{center}

\end{center}

\caption{The parameters $\bar{\Gamma}$~h$^{-1}$ (curve~1)
and $\epsilon$ (curve~2) versus $\Delta T$~$^{\circ}$C
for amorphous PET (A) and versus $f$\% for semicrystalline PET (B).
Symbols: treatment of observations \protect{\cite{AH82,DCM99}}.
Solid lines: approximation of the experimental data by Eqs. (9) and (11).
A: $b_{0}=3.686$, $b_{1}=0.154$, $c_{0}=-2.424$, $c_{1}=0.073$;
B: $b_{0}=0.618$, $b_{1}=0.038$, $c_{0}=0.247$, $c_{1}=0.071$}
\end{figure}

\begin{figure}[t]
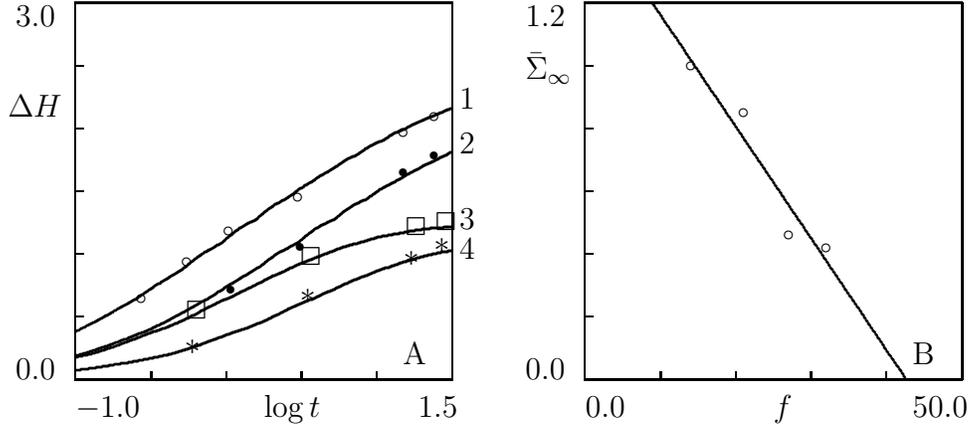

\begin{center}
%
\end{center}

\caption{A---The relaxation enthalpy $\Delta H$ J/g versus time $t$ h
for semicrystalline PET ($T_{\rm g}=80$~$^{\circ}$C)
quenched from $T_{0}=100$ to $T=65$~$^{\circ}$C.
Symbols: experimental data  \protect{\cite{DCM99}}.
Solid lines: prediction of the model with $\bar{\Sigma}_{0}=0.2$
and $\Lambda=1.5$.
Curve~1: $f=14$\%;
curve~2: $f=21$\%;
curve~3: $f=27$\%;
curve~4: $f=32$\%.
B---The parameter $\bar{\Sigma}_{\infty}$ versus $f$\%.
Circles: treatment of observations.
Solid line: approximation of the experimental data by Eq. (10)
with $a_{0}=1.519$ and $a_{1}=0.036$}
\end{figure}

\begin{figure}[t]
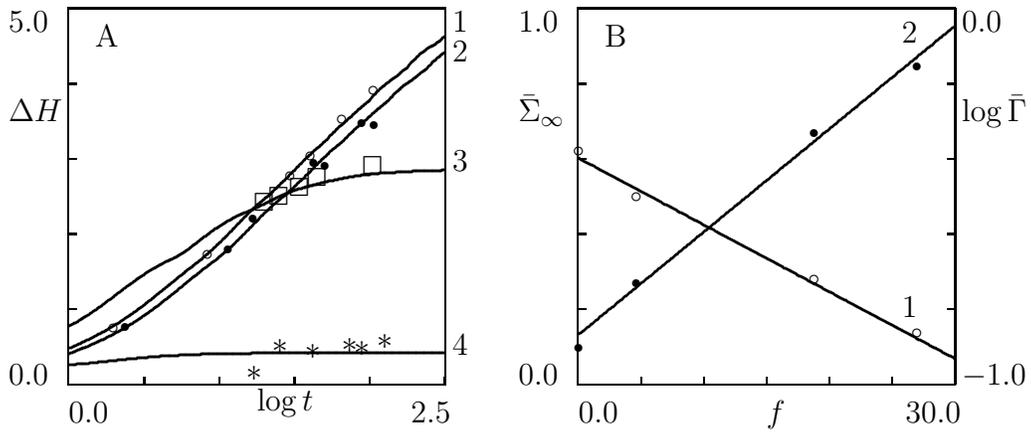

\begin{center}

\end{center}

\caption{A---The relaxation enthalpy $\Delta H$ J/g versus time $t$ days
for semicrystalline PET ($T_{\rm g}=77$~$^{\circ}$C)
uniaxially stretched at $T_{0}=100$~$^{\circ}$C
and quenched to $T=40$~$^{\circ}$C.
Symbols: experimental data \protect{\cite{MJ95}}.
Solid lines: prediction of the model with $\bar{\Sigma}_{0}=0.12$,
$\epsilon=0.2$ and $\Lambda=3.38$.
Curve~1: $\lambda=1.0$;
curve~2: $\lambda=2.0$;
curve~3: $\lambda=3.0$;
curve~4: $\lambda=4.0$.
B---The dimensionless parameter $\bar{\Sigma}_{\infty}$
(curve~1) and the parameter $\bar{\Gamma}$ days$^{-1}$ (curve~2)
versus $f$\%.
Circles: treatment of observations.
Solid lines: approximation of the experimental data by Eqs. (10) and (11)
with $a_{0}=0.603$, $a_{1}=0.018$ and $b_{0}=-0.868$, $b_{1}=0.027$}
\end{figure}
\end{document}